\documentclass[journal,twocolumn]{IEEEtran}
  \usepackage{graphicx,multirow,multicol,textcomp,booktabs}
  \usepackage[nospace,noadjust]{cite}

\begin{document}
%
\title{Wireless communication for safe UAVs:\\
From Long-Range Deconfliction to short-range collision avoidance
}
%
%
%

\author{Evgenii~Vinogradov,
        Franco~Minucci,
        and~Sofie~Pollin}

%
%

\markboth{IEEE Vehicular Technology Magazine Special Issue on Communications Support for Unmanned Air Transportation}%
{Shell \MakeLowercase{\textit{et al.}}: Bare Demo of IEEEtran.cls for IEEE Journals}
%



\maketitle

\begin{abstract}
Small drones are becoming a part of our everyday life. They are used in a wide variety of commercial applications, and the number of drones in the air is steadily growing. To ensure the safe operation of drones, traffic management rules must be designed and implemented by avionics and telecommunication experts. In this article, we propose to establish a common terminology for these two communities. We first describe the traffic management architecture and services. Next, we overview several approaches for defining the inter-drone separation distances ensuring the safe operation. Moreover, we analyze which existing technologies can be useful for each of these definitions. Finally, we present measurement results indicating that our new Wi-Fi-based messaging scheme is a potentially useful tool for the drone traffic management system.
\end{abstract}


%
\IEEEpeerreviewmaketitle

\section{Introduction}
%
%
%
%
\IEEEPARstart{U}{nmanned Aerial Vehicle} (UAV)-enabled solutions are becoming very popular. UAVs (or drones) are attractive owing to their flexibility and potential cost efficiency in comparison with conventional aircraft. While in some countries drones are perceived as “game changers” and “development enablers” \cite{ref:1}, in other areas, the public is rather concerned about safety and security issues aroused by the UAV-use.  Moreover, it is not fully understood how the wide-scale drones’ applications will influence conventional Air Traffic Management (ATM). 

National and supranational authorities (e.g., Federal Aviation Administration - FAA, European Union Aviation Safety Agency – EASA, International Civil Aviation Organization – ICAO) and industrial actors (Amazon, Google, DJI) are now developing systems for UAV Traffic Management (UTM). These services and products are vital for establishing trust between the authorities, the public, and industry. As it is anticipated that UTM and ATM systems will, at some point, coincide or overlap, the common terminology and approaches are vital. Even though drones are highly technological vehicles, their presence in the air, in the first step, will be regulated by safety rules of manned aviation which are the results of years of operational experience and technology maturation. Moreover, the authorities imposing air traffic rules have their inertia and reasonably give a higher priority to the manned aviation. We understand that this stage of UAV-development is an essential milestone in the process of making UAVs a global phenomenon and a potential development enabler. 

\subsection{UTM architecture and key services}
\begin{figure}[h!]
\centering
	\includegraphics[width=1\columnwidth]{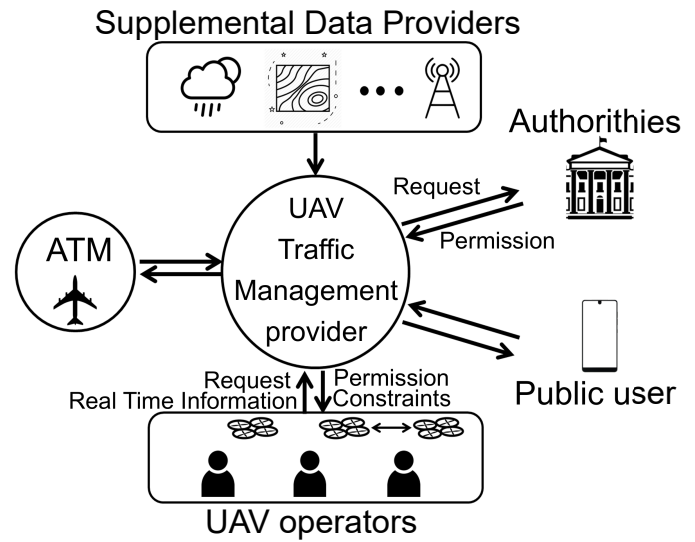}
	\caption{Schematic UTM architecture demonstrating the interactions between different air traffic agents}\label{fig:1}
\end{figure}

UTM is the all-encompassing framework for managing small UAVs, providing a set of services to the ATM system and UAV-operators. It includes everything concerning UAV operations: operation rules, registrations, waivers, and performance-based requirements \cite{ref:2}. The main aim of UTM is to achieve safe and efficient UAV operations. The UTM architecture (see Figure \ref{fig:1}) includes the following entities:
\begin{itemize}
    \item UAV operators,
    \item Regulators(FAA, EASA),
    \item Supplemental data providers (weather, terrain information, communication providers, etc.),
    \item Other stakeholders (e.g., public safety, the public).
\end{itemize}

UTM will provide multiple services targeting safe UAV operations (deconfliction support; a link between ATM and UAV operators; on-demand information to authorities and public; operational approval support of UAV operators).

Summarizing, UTM will allow the authorities and ATM providers to manage the shared airspace without being continuously involved in unmanned air traffic operations. At the same time, the regulator will have on-demand access to real-time operation status, UAV’s location, and flight plan, as well as a possibility of obtaining data for the post-hoc events-of-interest analysis.

In its turn, ATM will have to align several services with UTM \cite{ref:2}:
\begin{itemize}
    \item Airspace Organization and Management,
    \item Demand and Capacity Balancing,
    \item Airspace User Operations,
    \item Strategic Conflict Management, and
    \item Information Services.
\end{itemize}
We would like to emphasize that the principles and strategies for airspace reservations will have to include UAVs as new agents.

\subsection{Need of interdisciplinary terminology}
One of the essential parts of future UTM is wireless communication. The research community is very active in designing UTM and UAV-enabled wireless technologies, in general \cite{ref:3,ref:4,ref:5}. However, for creating technologies that have a higher chance to reach a practical implementation, the telecommunication research community must rely on knowledge, achievements, and requirements imposed by the avionic experts. We need to establish a common terminology to avoid confusion that is widely observed in literature dedicated to various wireless communication-related aspects of the UTM and UAV-enabled solutions. For instance, in telecommunications, “collision avoidance” is often used for describing techniques used to avoid resource contention during an information transmission. Moreover, the same term is used for describing a physical collision, as in \cite{ref:3,ref:5}, but this definition is still much more general than the one used in avionics.

In this article, we aim to establish a common terminology that can be used by both wireless communication and avionics experts. Moreover, we analyze the applicability of the existing wireless technologies in the light of these new definitions. The scope of this article is limited by one aspect of UTM, namely, by avoiding a physical collision of a UAV with other aerial vehicles and how various wireless technologies can help to achieve this goal. 

First, we give a clear definition of the conflict management (CM) and its layers (i.e., strategic deconfliction, remaining well-clear, and collision avoidance). Next, we describe how these layers are defined based on time or distance between UAVs. Finally, we indicate the existing technologies that can be used to accommodate each layers’ needs. Moreover, we show how a Wi-Fi-based solution can be used for deconfliction.


\section{Deconfliction vs. collision avoidance}

\begin{figure}[h!]
\centering
	\includegraphics[width=1\linewidth]{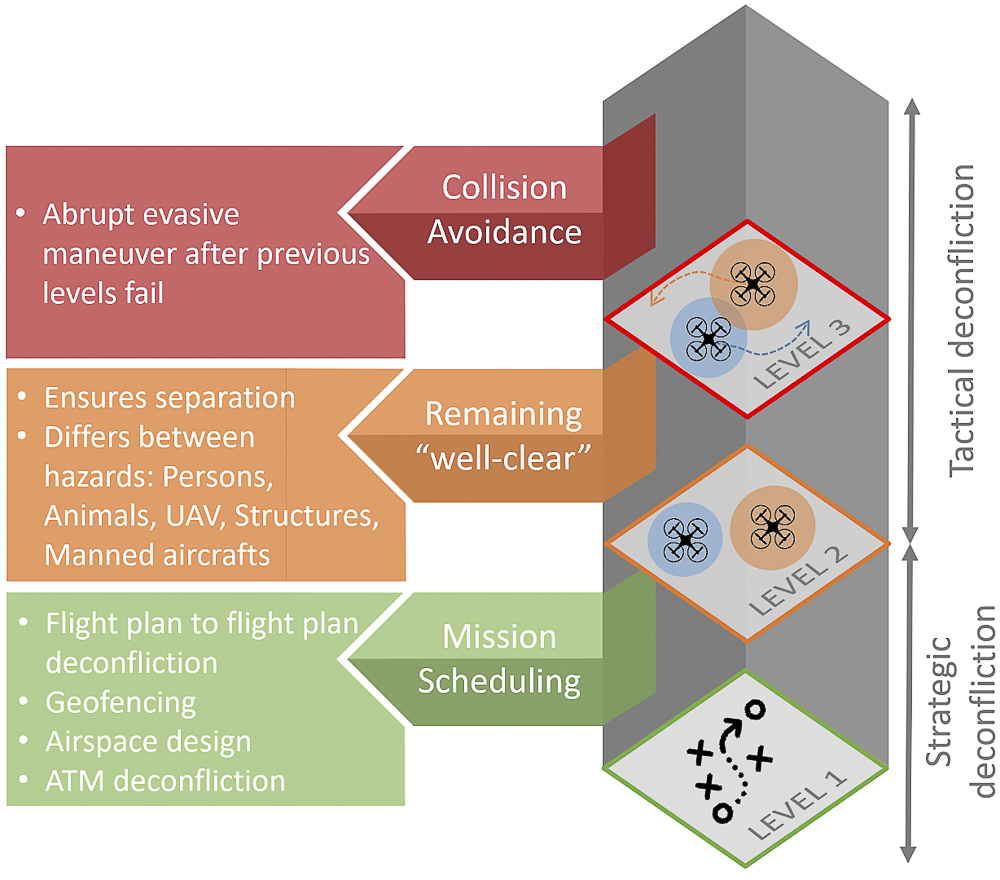}
	\caption{Conflict Management layers and corresponding procedures show that there is well-defined isolation between the layers}\label{fig:2}
\end{figure}

Whereas the definition of UTM is clear for telecommunication experts, we observe some confusion when collision avoidance is discussed. In avionics and the vast majority of regulatory documents, the process of ensuring that aerial vehicles do not physically collide is referred to as conflict management. Consequently, deconfliction must be performed to limit, to an acceptable level, the risk of collision between aircraft (including UAVs). CM consists of three independent layers that together ensure the minimization of collision probability (see Figure \ref{fig:2}). Note that each layer’s measures aim to reduce the need to apply the next layer’s procedures to an appropriate level as determined by UTM requirements. Consequently, the second and third layers will only be used when the higher layers cannot be used efficiently. Even though the layers are explicitly separated, it is recognized that a continuum exists from the earliest planning of the UAV activity through to the latest avoidance measure. Next, let us describe these layers as in \cite{ref:6}.
\subsubsection{Strategic deconfliction (or Mission scheduling)} is performed at the first layer of conflict management. It is achieved through the airspace organization and management (flight planning), demand and capacity balancing, and traffic synchronization components. Strategic actions will typically occur prior to departure. However, they are not limited to pre-departure, particularly in the case of longer duration flights.
\subsubsection{Separation provision} (more known as “remaining well-clear”) is the second layer of conflict management. It is the tactical process of keeping aircraft away from hazards by at least the appropriate separation minima. It is an iterative process consisting of: 
\begin{itemize}
    \item Conflict Detection: based on the current position of the aircraft involved and their predicted trajectories,
    \item Solution Formulation: selection of the separation modes to maintain separation of aircraft from the hazards within the appropriate conflict horizon; new trajectories should be checked to ensure that they are free from conflicts,
    \item Solution Implementation: trajectory modification,
    \item Monitoring of Solution Implementation: ensuring the appropriate separation minima. 
\end{itemize}
\subsubsection{Collision avoidance} is the third layer of conflict management and must activate when the separation mode has been compromised. Collision avoidance maneuver is the last resort to prevent an accident.

Note that layers 2 and 3 are referred to as \emph{Tactical deconfliction}. In the next section, we will summarize state-of-the-art approaches to define the volumes/boundaries of the three CM layers based on time and space separation of two UAVs. 

\section{Making collision avoidance Well-clear for everyone}
\subsection{Time-based definition}
For the manned aviation, the three aforementioned layers historically were defined using the expected time of the potential collision as a reference point. It is proposed to use the same approach for UTM as well \cite{ref:7}. In this case, the levels are defined using the following timing:
\begin{itemize}
    \item Level 1 (Strategic deconfliction): 24 hours – 2 minutes;
    \item Level 2 (Remaining well-clear): 2 minutes – 30 seconds;
    \item Level 3 (Collision avoidance): 30 seconds – 0 seconds.
\end{itemize}

\emph{Critique: }This definition is setting clear boundaries between the CM layers and makes the level definition very intuitive at first glance. Unfortunately, there are no guidelines on how to treat vertical separation, which is one of the ways to mitigate collisions. Additionally, when we try to map time-to-collision into distance and analyze the collision probability (for example, as in \cite{ref:8}), we face the fact that the layer boundaries blur due to the high range of UAVs velocities (from nearly 0 m/s to 74 m/s \cite{ref:9}). Moreover, there is no recommendation which speed must be used for the time-to-collision estimation (instantaneous, planned, cruise, or maximum). This ambiguity can result in significant over- or underestimation of the required distance separation. For example, let us assume a situation when two racing drones (e.g., RaceX\footnote{Guinness World Records, available online https://www.guinnessworldrecords.com/news/commercial/2017/7/the-drone-racing-league-builds-the-worlds-fastest-racing-drone-482701} ) fly towards each other. In this case, well clear range is 4.4 - 17.7 km, which looks exaggerated for an aircraft with the size and weight of a pigeon: the collision probability calculated as in \cite{ref:8} is negligibly small even at the 1 km distance due to the UAV’s dimensions.

\subsection{Distance-based definition}

\begin{figure}[h!]
\centering
	\includegraphics[width=1\linewidth]{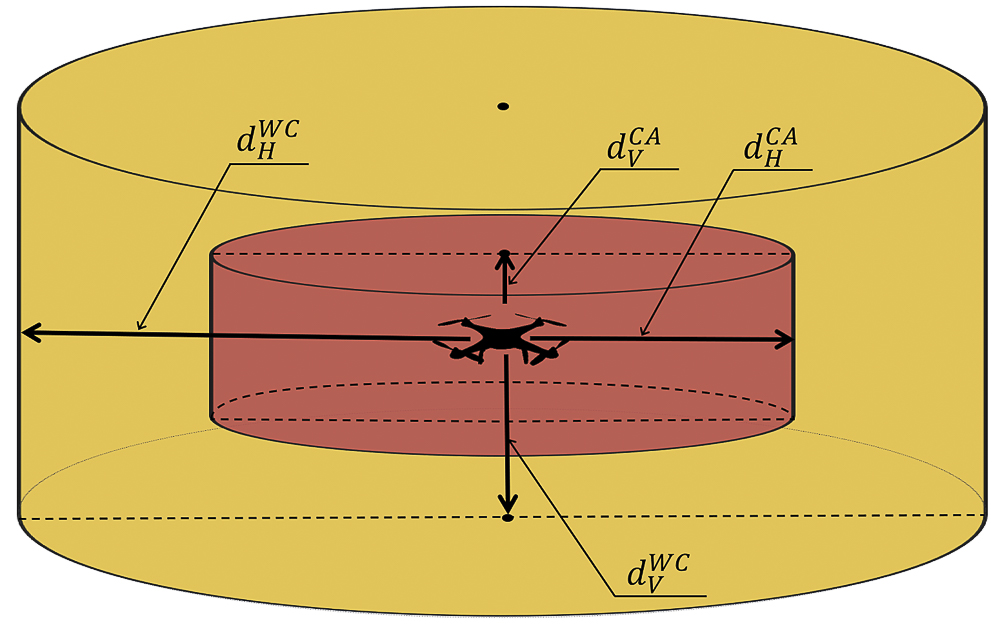}
	\caption{Distance-based “Hockey puck” well-clear (orange cylinder) and collision definition (red cylinder)}\label{fig:3}
\end{figure}

Another popular approach is the distance-based definition of the three CM volumes. The volumes corresponding to the Levels 2 and 3 are often represented by cylinders (so-called “hockey puck” – see Figure \ref{fig:3}). In the figure, the Well-Clear (WC) and Collision avoidance (CA) cylinder dimensions are denoted as $d_H$ and $d_V$ with the corresponding superscripts. Next, we describe two hockey puck recommendations applicable to sub-urban and urban environments.

In \cite{ref:9}, A. Weinert et al. proposed a recommendation for UAV systems (see Table 1). The authors thoroughly analyzed maximum and cruise airspeed of nearly 500 UAVs (fixed- and rotary-wing) as well as the UAVs dynamics depending on their missions and the vendor performance guidelines. As can be seen, the range of distances is large. However, for each WC hockey puck, the authors calculated the probability the level 2 procedures would fail. Logically, this probability is higher for smaller WC volumes. For example, if another drone is closer than 600 and 75 meters in horizontal and vertical domains, respectively, then there is 10\% probability that this drone will violate a smaller volume (150 by 30 meters). Consequently, both drones will have to actively and promptly avoid each other following the Level 3 procedures. Note that this is not a probability that the two UAVs will eventually collide.

\emph{Critique:} This definition is very appealing due to its simplicity. However, applying these conservative recommendations in an urban environment with a wide-scale drone deployment (delivery, surveillance, etc.) will be prohibitive for many UAV-assisted applications. We believe that this set of recommendations is an excellent first step for drone use-cases in rural and sub-urban areas.

\begin{table}
\centering
    \caption{Separation distances for the well-clear and collision avoidance volumes depending on the deployment environments}\label{tab:1}

    \begin{tabular}{c|c|c}
    \toprule
    &Horizontal &Vertical \\
    &separation, $d_H$&separation, $d_V$  \\
    \midrule
    \multicolumn{3}{c}{Sub-urban environment \cite{ref:9}} \\
    \midrule
     \multirow{2}{*}{Well-Clear}&600 - 1500 m& 75 - 90 m \\
     &(2000-3500 ft)&(250-300 ft)\\
    Collision avoidance&150 m (500 ft)&30 m (100 ft)\\
    \midrule
    \multicolumn{3}{c}{Urban environment \cite{ref:10}} \\
    \midrule
    Well-Clear&6 m (20 ft)& 7.3 m (24 ft)\\
    Collision avoidance&3 m (10 ft)&3.7 m (12 ft)\\
    \bottomrule
    \end{tabular}
\end{table}

For urban environments, another distance-based definition was proposed in \cite{ref:10}, where a much smaller WC and CA cylindrical volumes were defined (see Table 1). In this study, UAVs were moving between the buildings in several environmental settings. It was pointed out that the collision probability can be significantly reduced if the drone is able to sense objects potentially violating its well-clear volume in advance (6-8 seconds of the conflict horizon). The authors implemented a sophisticated deconfliction approach taking into account both static and mobile obstacles (buildings, UAVs, etc.) with a possibility to change the buffer regions and priorities.

\textit{Critique:} This definition is very liberal and can be applied in a dense urban environment. The resolution algorithm is quite complex, which can lead to high requirements for computational power if the number of conflicts is high. Further analysis is needed to assess the collision probability for a broader set of urban environments. 

Considering the layers definitions given above, it becomes evident that each layer of CM has its requirements for the used communication means and technologies.
\section{Wireless technologies for deconfliction}
Wireless communication plays a vital role in ATM and UTM. Needless to say that all communication with air traffic controllers is wireless. For example, one of the key technologies in manned aviation is Automatic Dependent Surveillance - Broadcast (ADS-B). It is a surveillance technology in which an aircraft determines its position via satellite navigation and broadcasts it every 500 ms. The information can be received by air traffic control ground stations and by other aircraft to provide situational awareness and allow self separation. Unfortunately, the widespread use of ADS-B is not considered feasible for UTM due to potential frequency congestion \cite{ref:11}. Note that in this work, we consider the distributed part of UTM where UAVs exchange their positions without an intermediate operator.

In this section, we indicate how the existing wireless technologies can be used at different CM layers. Moreover, we describe several ADS-B like solutions for UTM.
\subsection{Range and update rate}
An overview of the technologies is provided in Table 2. Note that the numbers given in the last column are based on UAV measurement campaigns performed in the referred papers. Lin et al. proposed using several technologies to send the UAV position in \cite{ref:12}. They recommend using LoRa (Long Range Wide Area Network), and APRS (Automatic Packet Reporting System). Work in \cite{ref:11} presents Reduced Power ADS-B (RP ADS-B), where the standard ADS-B transmit power (40W) was reduced by 20 dB to overcome the spectrum congestion issue. The measurements based on Wi-Fi Service Set Identifier (Wi-Fi SSID) and LoRa are described later in this article as practical use-cases. 

\begin{table}
\centering
    \caption{Wireless technologies applicable for the three layers of the conflict management. Strategic deconfliction requires the ground infrastructure. Tactical deconfliction relies on  coordinates broadcasting: the final choice depends on the targeted range and update requirements.}\label{tab:2}

    \begin{tabular}{c|c|c|c}
    \toprule
\multirow{2}{*}{Technology}& \multirow{2}{*}{Range}&\multicolumn{2}{c}{Update rate}\\
&&minimum&measured\\
\midrule
\multicolumn{4}{c}{Broadcasting}\\
\midrule
Bluetooth &	100 m&25 ms&\\
Bluetooth Low energy&50 m&25 ms&\\
ZigBee&100 m&25 ms&\\	
ANT&30 m&25 ms&\\
APRS \cite{ref:12}&20 km&5 s&11 - 33 s\\
ADS-B&370 km&0.5 s&\\
RP ADS-B \cite{ref:11}&1200 m&0.5 s&2 - 3 s\\
Wi-Fi SSID&800 m&60 ms&0.1 - 0.8 s\\
LoRa&15 km&5.16 s&5.16 - 30 s\\
\midrule
\multicolumn{4}{c}{Ground Infrastructure}\\
\midrule
Wi-Fi&500 m&&\\
LTE&1 km&&\\
ADS-B&370 km&&\\
LoRa&15 km&&\\
\bottomrule
		\end{tabular}
\end{table}
\begin{enumerate}
    \item \emph{Strategic deconfliction} (Level 1) is not posing hard requirements on delay, latency, throughput, etc. The most crucial metric is coverage which often implies the presence of the ground infrastructure. For Layer 1, we recommend using technologies offering reliable coverage in large areas (e.g., LTE or LoRa).
    \item \emph{Tactical deconfliction} (Levels 2 and 3) is a much more complex task. The choice of appropriate technologies for these two CM layers is not obvious due to the difference in the well-clear and collision avoidance volume definitions, as shown in the previous section. However, it is evident that preference should be given to the technologies allowing broadcasting since there is no time to establish a connection with all the drones around. 
\end{enumerate}

In some cases (e.g., in a city), the separation distance is minimal (see Table 1), so it becomes critical to update the UAVs’ positions as frequently as possible. Hence for an urban environment, we do not recommend considering APRS, (RP) ADS-B, and LoRa. On the other hand, these technologies are suitable for ensuring the well-clear separation for sub-urban environments. Note that Wi-Fi SSID demonstrates the combination of the range and update rate which makes it the right candidate for Levels 2 and 3 CM in suburban environments.

\subsection{Non-collaborative deconfliction}

\begin{figure}
\centering
	\includegraphics[width=1\linewidth]{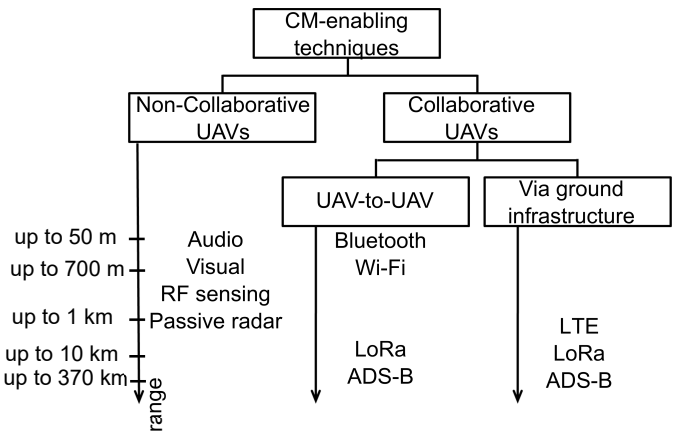}
	\caption{Classification of the wireless technologies for Conflict Management}\label{fig:4}
\end{figure}

The approaches listed in Table 2 based on the assumption that all drones have the same set of equipment and collaborate. First, it might not be the case in real life. Second, it is necessary to have a backup solution that can increase the CM reliability. Passive sensing can solve these two problems. In \cite{ref:13,ref:14}, analysis of several signals of different nature (Radiofrequency, audio, and visual) resulted in a conclusion that a UAV-mounted detection equipment is more effective than the terrestrial sensors. Various wireless technologies and techniques applicable for the CM purposes are summarized in Figure \ref{fig:4}. Note that in the figure, the maximum distance to the closest Base Station (BS) is shown for LoRa and LTE. However, the information can propagate farther to the coverage area of another BS theoretically providing global coverage. Similarly, a network of sensors can cover a larger area for detecting non-collaborative UAVs.

\subsection{Case Study 1: ADS-B like messaging via Wi-Fi SSID}
In \cite{ref:3}, it was shown that Wi-Fi suffers from interference, which results in problems with establishing a reliable connection between 2 nodes (2 UAVs, or UAV and the ground controller/UTM infrastructure). However, embedding the coordinates and drone ID in Wi-Fi SSID allows to broadcast this critical data in an ADS-B like manner, as it was shown in \cite{ref:15}. In other words, the coordinates can be encoded into the Wi-Fi network name that can be read by any UAV equipped with a Wi-Fi module. The advantages of this approach are that i) there is no need to establish a connection to Wi-Fi Access Point (AP) and ii) UAVs can exchange coordinates directly, without involving the ground infrastructure.

Since the majority of inexpensive and lightweight Wi-Fi modules have only one RF chain, the operations of SSID scanning and broadcasting cannot be performed at the same time. In \cite{ref:15}, it was shown that the most efficient scheme is when Wi-Fi module scans for messages at a fixed channel and broadcast its messages at all available channels. These two states (i.e., scan and broadcast) alternate in a random manner. Even though usually the states have different duration due to the hardware limitations, it is ensured that time is equally shared between both operations (50\% in this article) to maximize the probability of the successful reception. Note that the presented scheme does not require synchronization.
\begin{figure}
\centering
	\includegraphics[width=1\linewidth]{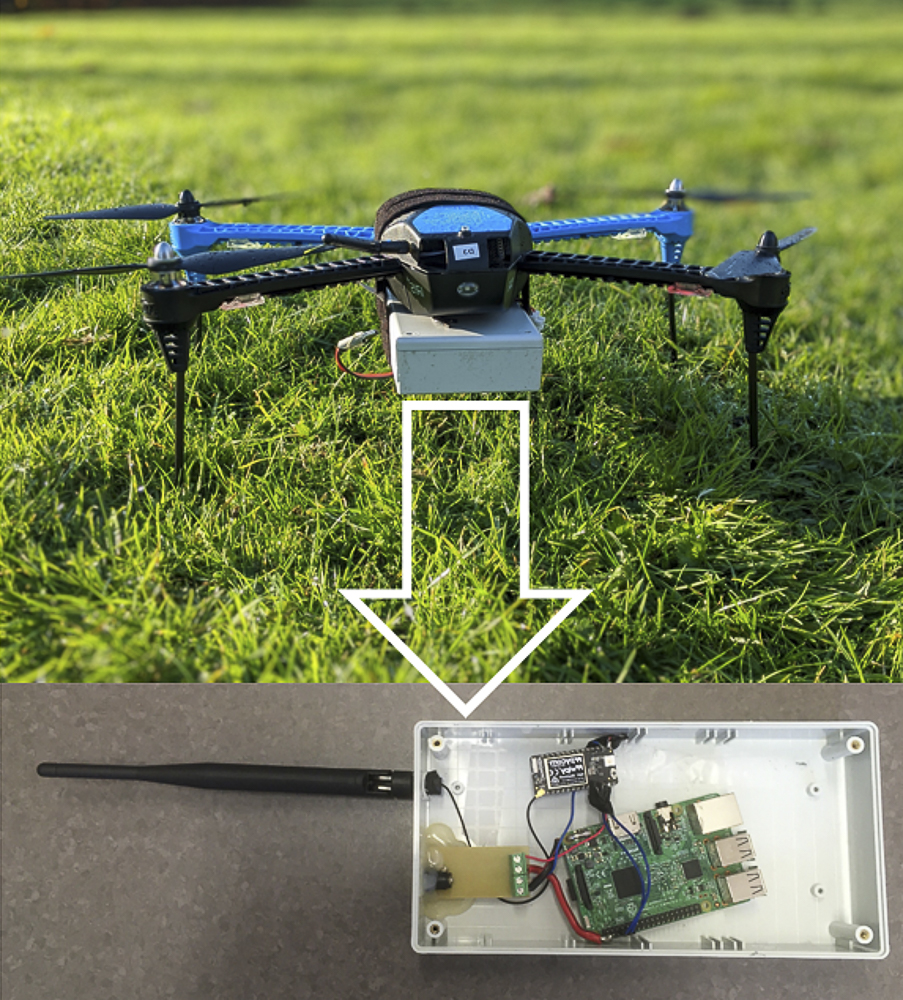}
	\caption{Prototype of our Wi-Fi and LoRa location broadcast system, emulating ADS-B for drones }\label{fig:5}
\end{figure}

\begin{figure}
\centering
	\includegraphics[width=1\linewidth]{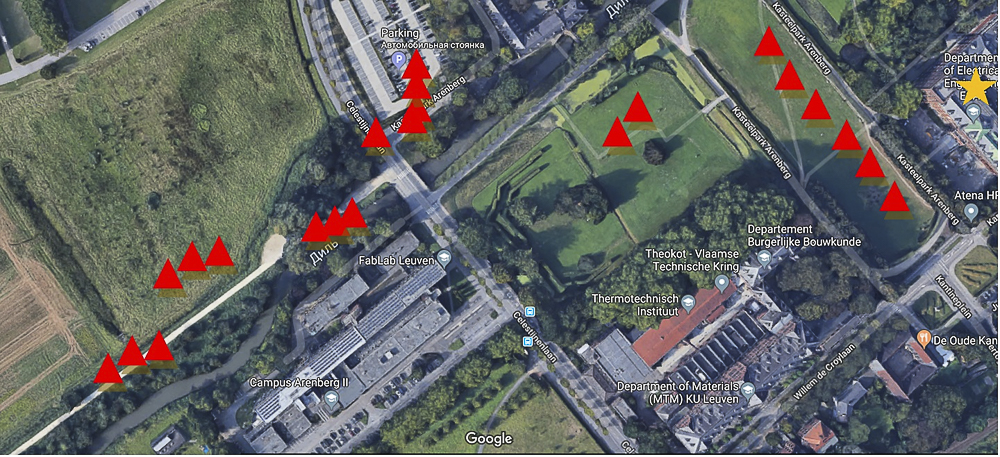}
	\caption{Map of the Wi-Fi SSID measurement campaign: star and triangles denote the nodes’ positions}\label{fig:6}
\end{figure}

The described idea was implemented in practice (see Figure \ref{fig:5}). One communication node was attached to a UAV, and another one was deployed on the top of a tall building imitating another drone (absence of objects in the proximity was ensured). Two Wi-Fi modules (ESP32 working in 2.4 GHz band) were reporting the received coordinates to the computers (Raspberry Pi). The flight area in Heverlee, Belgium, is demonstrated in Figure \ref{fig:6} (maximum distance between two Wi-Fi modules is around 700 m).

\begin{figure}
\centering
	\includegraphics[width=1\linewidth]{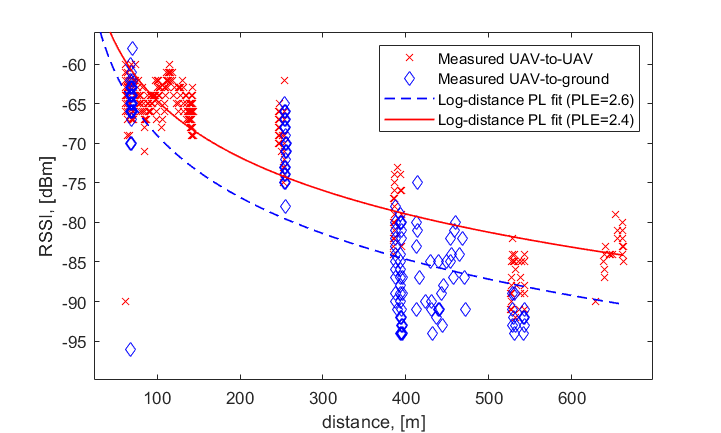}
	\caption{Distance-dependent Received signal strength. UAVs experience less signal attenuation in comparison with the ground users}\label{fig:7}
\end{figure}

Figure \ref{fig:7} compares Received Signal Strength Indicators (RSSI) for UAV-to-UAV and UAV-to-ground scenarios. Moreover, Log-distance Path Loss (PL) model parameters were estimated for these cases using the linear regression in Matlab. It turns out that the signal experiences less attenuation in the air due to fewer obstructions: Path Loss Exponent (PLE) equals 2.6 and 2.4 for the ground and hovering levels, respectively. This finding is in line with results in \cite{ref:3} and proves that Wi-Fi can be used for communicating with UAVs at longer distances corresponding to the well-clear range (e.g., on the ground there is no reception at more than 550 m, whereas in the air we can receive at almost 700 m).

\begin{figure}
\centering
	\includegraphics[width=1\linewidth]{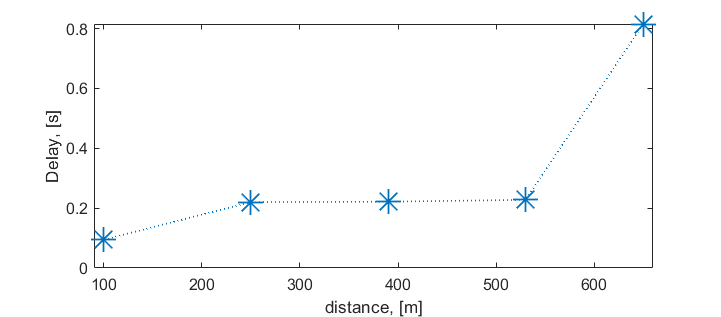}
	\caption{Distance-dependent delay between two received messages. The proposed scheme outperforms RP ADS-B, where the delay was 2-3 s [11]}\label{fig:8}
\end{figure}

Note that RSSI can be extracted only from successfully received messages. This means that even though RSSI is a meaningful metric, it does not provide a full picture. For a complete analysis, we estimated how frequently UAV could receive messages (updates of the other drone location). First, we grouped the received messages into five clusters with different distances and calculated the mean delay (see Figure \ref{fig:8}). On average, the described scheme allows us to get position updates every 95 ms when the drones are separated by 100 m distance. This is enough even for collision avoidance purposes (it corresponds to 0.95 m of traveled distance for 10 m/s - an average cruise speed of rotary UAVs reported in \cite{ref:9}). The delay between updates grows as the inter-UAV distance increases (see Figure \ref{fig:8}) due to the channel influence. 

Interestingly, RSSI levels at the farthest points are better than at 530 m, but in Figure \ref{fig:8}, we observe that many messages are lost, which is resulting in a slower update rate. This is explained by the fact that the UAV received a strong signal when the nodes were in line-of-sight (LOS); however, the messages that were lost did not contribute to RSSI.

\subsection{Case Study 2: ADS-B like messaging via LoRa}
LoRa is the technology that allows several types of communication: i) between two LoRa modules and ii) with a set of BSs. Both modes can be useful for UTM: strategic deconfliction can be performed via ground infrastructure since it is a pre-flight procedure and the peer-to-peer functionality (to avoid the delay introduced by the LoRa network) can be used for direct UAV-to-UAV communication in the tactical deconfliction layers. Here we consider only peer-to-peer scenario. The same set of equipment was used. The only difference that instead of Wi-Fi, we enabled LoRa communication with FiPy modules (by Pycom). We used Spreading Factor SF=7 because it results in a range sufficient for the tactical deconfliction.

In this measurement campaign, we aim to check how often we can receive the location updates via the LoRa-based communication. Due to the duty cycle requirements for LoRa, two consecutive messages must be separated by at least 5 seconds. 

We performed several flights; the distances between the nodes were 1 and 2 km. At 1 km, all messages were successfully received. For longer distances, we observe the messages' loss. At 2 km, only 20\% of messages received, which results in 25-30 seconds delay between two location updates. The delay is not critical for such long distances. Moreover, it decreases as the UAVs getting closer to each other due to the better channel \cite{ref:3} and, consequently, a higher probability of successful message delivery. However, we would like to underline that LoRa can be used only for the first and second layers of CM.
\section{Challenges and Future Research Trends}
\subsection{Location exchange and command and control via ground infrastructure}
In prior art, cellular-connected drones received a lot of attention \cite{ref:3,ref:5}. However, most of the published works are concentrated on the link-level metrics. We have a good understanding of the altitude-variant interference behavior, coverage, and other similar metrics. However, the end-to-end performance estimation is missing in literature. In UTM, the end-to-end (i.e., UAV-to-UAV) latency becomes a critical metric that must be studied.  
\subsection{Non-collaborative UAV detection and localization}
The reliability of future UTM services will depend on the ability to detect both collaborative and non-collaborative UAVs in the airspace. Several techniques have been proposed in the literature. However, an analysis of the required sensors’ density and reliability is necessary as well as a thorough comparison of all mentioned techniques (i.e., passive audio and RF sensing, passive radars, cameras) from the economic, energy efficiency, and deployment complexity points of view.
\subsection{New communication technologies for UAV}
In this article, we described how the existing technologies could accommodate the conflict management needs. One important extension is to explore the potential of new techniques (e.g., Massive Multiple-Input-Multiple-Output, millimeter waves, URLLC) for communication with UAVs. The new applications have already received some attention \cite{ref:3,ref:5}. However, they will require further analysis, simulations, and field measurements

\section{Conclusion}
The shipments of amateur and small commercial UAVs have skyrocketed in the last few years. This results in a challenge of integrating new agents to the shared airspace. UTM is seen as a tool to make the drones objectively safer as well as to demonstrate these increased safety standards to the public. In this article, we focused on establishing a common terminology for avionics and telecommunication experts, resulting in a better understanding of the needs and requirements of the two research communities. We summarized the state of the art in the conflict management layers definition. Moreover, we analyzed the layers from the wireless connectivity point of view based on a real-life UAV measurement campaign. We indicated how to ensure the necessary communication quality by using existing technologies (e.g., LTE and LoRa are suitable for the strategic deconfliction, whereas LoRa and Wi-fi can be used for remaining well-clear and the collision avoidance). Summarizing, we found that existing LoRa and Wi-Fi modules targeting terrestrial usage could support the initial deployment of UTM for low-altitude drones.

\section*{Acknowledgements}
This work is part of a project that has received funding from the SESAR Joint Undertaking (JU) under grant agreement No. 763702. The JU receives support from the European Unions Horizon 2020 research and innovation programme and the SESAR JU members other than the Union.

\ifCLASSOPTIONcaptionsoff
  \newpage
\fi



%

%

\begin{IEEEbiographynophoto}{Evgenii Vinogradov}
received the Dipl. Engineer degree in Radio Engineering and Telecommunications from Saint-Petersburg Electro-technical University (Russia), in 2009. After several years of working in the field of mobile communications, he joined UCL (Belgium) in 2013, where he obtained his Ph.D. degree in 2017. His doctoral research interests focused on multidimensional radio propagation channel modeling. In 2017, Evgenii joined the electrical engineering department at KU Leuven (Belgium) where he is working on wireless communications with UAVs and UAV detection.
\end{IEEEbiographynophoto}

\begin{IEEEbiographynophoto}{Franco Minucci}
received his BSc and MSc in electronics engineering from the University of Calabria in Italy. After graduation, he worked for three years in Italy mostly as hardware designer for telecom equipment. In 2011 he moved to Belgium where he worked for Ansem NV, IMEC vzw and Nokia (former Alcatel Lucent) in Antwerp. Currently, he is a PhD Student in KU Leuven where he is working on a wireless system for automatic air conflict management and collision avoidance.
\end{IEEEbiographynophoto}

\begin{IEEEbiographynophoto}{Sofie Pollin}
received the Ph.D. degree (Hons.) from KU Leuven in 2006. From 2006 to 2008, she continued her research on wireless communications, energy-efficient networks, cross-layer design, coexistence, and cognitive radio at UC Berkeley. In 2008, she returned to imec to become a Principal Scientist at the Green Radio Team. She is currently an Associate Professor with the Electrical Engineering Department, KU Leuven. Her research centers around networked systems that require networks that are ever more dense, heterogeneous, battery powered, and spectrum constrained. She is a BAEF Fellow and a Marie Curie Fellow.
\end{IEEEbiographynophoto}

\end{document}